%% file: paper.tex
\documentclass{sig-alternate-05-2015}
\usepackage{url}

\usepackage{amssymb,amsmath,amsfonts}
\usepackage{graphicx}
\usepackage{times}
 \usepackage{ifpdf}
\usepackage{algorithm}
\usepackage[algo2e, ruled, vlined]{algorithm2e}
\usepackage{caption}
\usepackage{subcaption}
\usepackage{authblk}
\usepackage{hhline}
\usepackage[inline]{enumitem}
\usepackage{framed}
\usepackage{array}

\newcommand{\ignore}[1]{}

\newcommand{\onlyinproc}[1]{}

\newtheorem{thm}{Theorem}[section]
\newtheorem{theorem}{Theorem}[section]

\newtheorem{lemma}[thm]{Lemma}

\SetKwFunction{Merge}{Merge}
\SetKwFunction{Initialize}{Init}
\SetKwFunction{CardEst}{CardEst}

\newcommand{\dWS}{dwsHH }
\newcommand{\cWS}{cwsHH }
\newcommand{\idWS}{Integrated dwsHH }

\newcommand{\dWSnoSp}{dwsHH}
\newcommand{\cWSnoSp}{cwsHH}
\newcommand{\idWSnoSp}{Integrated dwsHH}

\begin{document}

\title{Efficient Distinct Heavy Hitters for DNS DDoS Attack Detection}



\author{}
\author[1]{Yehuda Afek}
\author[2]{Anat Bremler-Barr}
\author[1]{Edith Cohen}
\author[1]{Shir Landau Feibish}
\author[1]{Michal Shagam}
\affil[1]{Blavatnik School of Computer Sciences, Tel-Aviv University, Israel}
\affil[2]{Computer Science Dept., Interdisciplinary Center, Herzliya, Israel}
\affil[ ]{\textit{afek@cs.tau.ac.il, bremler@idc.ac.il, edith@cohenwang.com, shirl11@post.tau.ac.il, michalshagam@mail.tau.ac.il}}

\maketitle {\let\thefootnote\relax\footnote{This research was
supported by a grant from the Blavatnik Cyber Security Councile, by
a grant of the European Research Council (ERC) Starting Grant no.
$259085$, and by the Ministry of Science and Technology, Israel.}

\begin{abstract}
\input{abstract}

\end{abstract}


\pagenumbering{arabic}
\setcounter{page}{1}

\input{introductionRevised}

\input{Edith}






\input{evaluation}

\input{newDNS}

\input{applications}

\input{relatedWork}

\section{Conclusion and Future Work}
\label{sec:conc} We have presented new and efficient algorithms for
distinct Heavy Hitters and combined Heavy Hitters
detection, as well as a system for detection of RSDDoS attacks
on the DNS service.

We are currently working on a more robust DNS random attack detection and mitigation system for both non-existent domain attacks and other forms of random subdomain attacks.
{\small
\bibliographystyle{plain}
\bibliography{cycle}
}

\end{document}

%% file: abstract.tex


Motivated by a recent new type  of randomized Distributed Denial of Service (DDoS) attacks on the Domain Name Service (DNS),
we develop novel and efficient distinct heavy hitters algorithms and build an attack identification system that uses our algorithms.

Heavy hitter detection in streams is a fundamental problem
with many applications, including detecting certain DDoS
attacks and anomalies. A (classic) heavy hitter (HH) in a stream
of elements is a key (e.g., the domain of a query) which appears in many elements (e.g., requests).
When stream elements consist of a $\langle key, subkey \rangle$ pairs, ($\langle domain, subdomain \rangle$) a distinct heavy hitter (dhh) is a key that is paired with a large number of different
subkeys. Our dHH algorithms are considerably more practical than
previous algorithms. Specifically the new fixed-size algorithms are simple to
code and with asymptotically optimal space accuracy tradeoffs.

In addition we introduce a new measure, a combined heavy hitter (cHH), which is a key with a large
combination of distinct and classic weights. Efficient algorithms are also presented for cHH detection.

Finally, we perform extensive experimental evaluation on real DNS attack traces, demonstrating
the effectiveness of both our algorithms and our DNS malicious queries identification system.

%% file: introductionRevised.tex

\section{Introduction}
\label{Introduction:sec}

\def\SH{{\textsf S\&H}}
\def\SHSpace{{\textsf S\&H }}

The Domain Name System (DNS) service is one of the core services in the internet functionality.
Distributed Denial of Service (DDoS) attacks  on  DNS  service  typically  consist  of  many queries  coming  from  a  large  botnet.  These  queries  are  sent to the root name server or an authoritative name server along the  domain  chain.  The  targeted  name  server  receives  a high volume of requests, which may degrade its performance or disable it completely. Such attacks may also contain spoofed source addresses resulting in a reflection of the attack or  may  send  requests that generate large responses (such as an ANY request) to use the DNS for amplification attacks.

According to Akamai's state of the internet report~\cite{AkamaiQ12016} nearly $20\%$ of DDoS attacks in $Q1$ of $2016$ involved the DNS service. Moreover, even some of the Internet's DNS \emph{root} name servers were targeted~\cite{VERISIGN}. 

One type of particularly hard to mitigate DDoS attacks 
are randomized attacks on the DNS service. In these attacks, queries for many different non-existent
subdomains (subkeys) of the same  primary domain (key) are
issued~\cite{NonsenseName:Liu2015}).  Since the result of a query to a
new subdomain is not
cached at the DNS resolver, these queries are propagated to the domain authoritative server, overloading both these servers and the
open resolvers of the Internet Service Provider.

Motivated by this DNS attack vector we develop in this paper a new distinct heavy hitter algorithm which is more practical and accurate than previous algorithms, and build a system for the mitigation of this attack based on this algorithm.

Detecting heavy hitters in a stream of requests is a fundamental problem
in data analysis, with applications that include monitoring for
malicious activities and other anomalies.
Consider a stream of DNS queries, with the top-level domain serving as the key. A key that appears a large number of
times in the query stream constitutes a {\em ``classic'' heavy hitter} (e.g., google.com, cnn.com, etc,).
In addition, each query's sub-domain serves as the subkey (e.g., mail., home., game1., etc,), and a key with many different subkeys is then a {\em distinct heavy hitters (dHH)}.
As another example, in the case of a spoofed TCP SYN flood DDoS attack, the packets received consist of the same destination and many different
sources, and therefore this attack may be represented as a dHH as opposed to a classic heavy hitter where most of the traffic could be due to a
small number of different sources. Finally, a {\em combined} heavy hitter (cHH), a notion we present and
motivate here, is a key with a large combination of distinct and
classic weights. Intuitively, a cHH is a key that combines both a ``classic''
heavy hitter as well as a distinct heavy hitter, meaning it both
appears a large number of times in the stream and has a high number
of different sub-keys.

\subsection{Contributions of this Paper}

\subsubsection{Algorithms}

Our main contributions are novel practical sampling-based structures for {\em distinct heavy hitter (dHH)} and {\em combined heavy hitter} (cHH) detection whose size (memory requirements in number of cache entries) are only $O(\epsilon^{-1})$ keys, where every key with weight of at least $\epsilon$ fraction of the (respective) total weight is detected with high probability. The total weights for HH, dHH and cHH are respectively the total number of items, the total number of distinct (key;subkey) pairs, and a weighted sum of the two.

Our proposed fixed-size dHH algorithm, named {\em Distinct Weighted Sampling (\dWSnoSp)}, requires a constant amount of memory as
opposed to the well known Superspreaders solution
\cite{superspreaders:NDSS2005} which uses a linear, to the input stream length, amount of memory.
Moreover, our use of sampling-based distinct counters is a significant
practical improvement over Locher's relatively new fixed-size solution
\cite{Locher:spaa2011} which utilizes linear-sketch based distinct
counters, which are much less efficient in practice. In addition, our
dHH algorithm produces a cardinality estimate for each key. This estimate is of much higher accuracy than the estimate
produced by Locher, while the Superspreaders do not provide comparable
estimates.

Our fixed-size cHH algorithm, named {\em  Combined Weighted Sampling (\cWSnoSp)} is, to the best of our knowledge, the first constant memory algorithm proposed which considers both the key volume as well as the distinctness of its subkeys.

Generally, approximate distinct heavy hitters algorithms exhibit a tradeoff between detection accuracy and the amount of space they require. Cardinality estimate accuracy is even more difficult to achieve with a fix-size structure since a key may be evicted from the cache and then re-enter the cache which presents some uncertainty with regards to cardinality. We provide a solution using a fix-size structure which outperforms known solutions both in terms of cardinality accuracy and practicality.
A more detailed comparison of our results to previous work is
shown in Section~\ref{related:sec}.


\ignore{
Our cHH structure significantly improves over
 the naive approach of maintaining separate structures
 for HH and dHH: The latter has overhead due to the
 overlap in the sets of cached keys in the two structures,
and also requires much larger structure sizes.
}

\subsubsection{Applications}

We design a system to detect randomized DNS request attacks.
The design uses our \dWS and \cWS algorithms and is enhanced to be a complete system for this specific
application. Additional applications of our \dWS and \cWS algorithms 
are given in Section \ref{dns:sec}




\subsubsection{Evaluation}
We demonstrate, via experimental evaluations on both real internet traces and synthetically generated data, the effectiveness of
our \dWS and \cWS algorithms as well as our system for detection of randomized DNS request attacks.

\subsection{Paper Organization}

Section~\ref{prelim:sec} provides required definitions and background. In Section~\ref{distinctSH:sec} we discuss our algorithms for distinct weighted sampling, followed by a discussion of our combined weighted sampling scheme in Section~\ref{combo:sec}. Section~\ref{eval:sec} describes experimental evaluation of our algorithms.
In Section~\ref{dns:sec} we describe the randomized DDoS attacks on DNS servers and our proposed system for mitigation of such attacks as well as an evaluation of our system.
Section~\ref{related:sec} summarizes related work, and Section~\ref{sec:conc} concludes the paper.

%% file: Edith.tex
\section{Preliminaries and Notations} \label{prelim:sec}
\SetKwArray{Counters}{dCounters}
\SetKwFunction{Merge}{Merge}
\SetKwFunction{Initialize}{Init}
\SetKwFunction{CardEst}{CardEst}
\SetKwFunction{rand}{rand}
\SetKwFunction{Hash}{Hash}
\SetKwFunction{seed}{seed}

\ignore{
 We provide more formal definitions and then
review two important ingredients in our approach:
Classic Sample and Hold stream sampling structures and
  approximate distinct counting structures. The notations used throughout the paper are summarized in Table~\ref{table:symbols}.
}

\subsection{Problem Definitions}

Formally, our input is modeled as a stream of elements, where each element has
 a primary key $x$ from a domain ${\cal X}$ and  a subkey $y$ from domain
$D_x$.  For each key, the {\em (classic) weight} $h_x$ is
the number of elements with key
  $x$, the {\em distinct weight}  $w_x$ is the number of  different subkeys in
  elements with key $x$, and, for a parameter $\rho\ll 1$, the {\em
    combined weight} is $b^{(\rho)}_x
  \equiv \rho h_x + w_x$.
In the particular example
of a DNS resolver, $h_x$ is the total number of requests for a primary domain
$x$ and $w_x \leq h_x$ is the number of distinct subdomains.
Combined weights are interesting as they can be a more accurate
measure of the load due to key $x$ than one of $h_x$ or $w_x$ in
isolation:
All $h_x$ requests are processed
but the $w_x$ distinct ones are costlier.  For
DNS resolvers, the distinct requests are costlier because their responses aren't in the cache.

A key $x$ with weight that is at least an $\epsilon$
fraction of the (respective) total is referred to as a heavy hitter: When
$h_x \geq  \epsilon \sum_y h_y$, $x$ is a (classic) {\em heavy
  hitter (HH)}, when $w_x \geq \epsilon \sum_y w_y$, $x$ is a
 {\em distinct
  heavy hitter (dHH)} or {\em superspreader}
\cite{superspreaders:NDSS2005}, and when
$b^{(\rho)}_x \geq \epsilon \sum_y b^{(\rho)}_y$, $x$ is a
  {\em combined heavy hitter} (cHH).

The notations used throughout the paper are summarized in Table~\ref{table:symbols}.

\begin{table}[h]
  \begin{center}
    \begin{tabular}{|c||l|}
    \hline
Symbol & Meaning  \\ \hhline{|=||=|}
$x$ & key \\
$y$ & subkey \\
$h_x$ & number of elements with key $x$ \\
$w_x$ & number of different subkeys in elements with key $x$ \\
$m$ & $\max_x w_x$ \\
$\tau$ & detection threshold \\
$k$ & cache size \\
$\ell$ & number of buckets \\
$\rho$ & combined weight parameter \\ \hline
\end{tabular}
\end{center}
\caption{Notations \label{table:symbols}}
    \end{table}

\subsection{Background}
\subsubsection{Sample and Hold} \label{SHsubsec}

The Sample and Hold (\SH)
algorithm   \cite{GM:sigmod98,EV:ATAP02} is
applied to a stream of elements, where each element has a key
$x$.  The  weight $h_x$ of a key $x$ is the number of elements with this key.

 The {\em fixed threshold} design is specified for a threshold $\tau$. The
 algorithm maintains a cache $S$ of keys, which is initially empty, and a
counter $c_x$ for each cached key $x$.
A new element with key $x$ is processed as follows:
If $x\in S$ is in the cache, the counter $c_x$ is incremented.  Otherwise, a counter $c_x \gets 1$ is initialized with probability $\tau$.
 The  {\em fixed-size}  design is specified for a fixed sample (cache)
 size  $k$ and works by effectively lowering the threshold
 $\tau$ to the value that would have resulted in $k$ cached keys.

An important property of \SH\ is that
the set of sampled keys is a
probability proportional to size without replacement (ppswor) sample
 of keys according to weights $h_x$ \cite{Rosen1972:successive}. 

\subsubsection{Approximate Distinct Counters}
A distinct counter is an algorithm that
maintains the number of different keys in a stream of elements.
 An exact distinct counter requires
state that is proportional to the number of different keys in the
stream. Fortunately, there are many existing designs and
implementations of {\em approximate}
distinct counters that have a small relative error but use state size
that is only logarithmic or double
logarithmic in the number of distinct elements
\cite{FlajoletMartin85,ECohen6f,BJKST:random02,hyperloglog:2007,ECohenADS:TKDE2015}.
The basic idea is elegant and simple:  We
apply a random hash function to
each element,  and retain the smallest hash value.  We can
see that this value, in expectation, would be smaller when there are
more distinct elements, and thus can be used to estimate this number.
The different proposed structures have different ways
of enhancing this approach to control the error. The tradeoff of
structure size and error are controlled by a parameter $\ell$: A
structure of size proportional to $\ell$ has normalized root mean square
error (NRMSE) of $1/\sqrt{\ell}$.
In Section \ref{distinctSH:sec} we
use distinct counters as a black box in our dHH structures, abstracted as
a class of objects that support the following operations:
\begin{itemize}
\item
$\Initialize$:  Initializes a sketch of an empty set
\item
$\Merge{x}$:  merge the string $x$ into the set ($x$ could already be a member of the set or a new string).
\item
$\CardEst$:  return an estimate on the cardinality of the set (with a
confidence interval)
\end{itemize}
 In Section \ref{integrate:sec}, we  also propose a
design where a particular algorithm for approximate distinct counting is integrated in
the dHH detection structure.

\section{Distinct Weighted Sampling} \label{distinctSH:sec}

 We now present our distinct weighted sampling schemes, which take as input elements that are key
  and subkey pairs.
We build on the fixed-threshold and fixed-size classic \SH\ schemes but
make some critical adjustments: First, we
apply hashing so that we can sample the distinct stream instead
of the classic stream. Second,
instead of using simple counters $c_x$ for cached keys as in classic \SH, we use approximate
  distinct counters applied to subkeys.  Third,
we maintain state per key that is suitable for estimating the weight
of heavy cached keys (whereas
classic \SH\ was designed for unbiased domain queries).

Our algorithms, in essence, compute heavy hitters using
weighted sampling.
A sample set of the keys is maintained during the execution of each of the algorithms (HH, dHH, or cHH).
The sample set constitutes a weighted sample
according to the respective counts so that the heavier keys, in
particular the heavy hitters, are much more likely to be included
than other keys. The counts in each of the algorithms are different; number of repetitions, measure of distinctness, and a combined measure, respectively. The algorithms maintain counts with each cached key which allow to produce the cardinality estimate for each output key.
\subsection{Fixed-threshold Distinct Heavy Hitters} \label{fixedtau:sec}

Our fixed-threshold distinct heavy hitters algorithm is applied with respect to a specified
threshold parameter $\tau$.  We make use of a random hash function $\Hash \sim U[0,1]$.   An element $(x,y)$ is processed as follows.
If the key $x$ is not cached, then if $\Hash(x,y)$  (applied to the key and subkey pair $(x,y)$) is below $\tau$, we initialize a $\Counters{x}$ object (and say that now $x$ is cached) and insert the string
$(x,y)$.   If the key $x$ is already in the cache, we merge the string $(x,y)$ into the distinct counter $\Counters{x}$. The pseudo code 
is omitted due to lack of space and can be found in the technical report~\cite{CompletePaper}.

\ignore{
\begin{algorithm2e}[h]
\caption{{\small Fixed-threshold Distinct Heavy Hitters}\label{distinctSHth:alg}}
\DontPrintSemicolon
{\small
\SetKwArray{Counters}{dCounters}
\SetKwFunction{Merge}{Merge}
\SetKwFunction{Initialize}{Init}
\SetKwFunction{CardEst}{CardEst}
\SetKwFunction{rand}{rand}
\SetKwFunction{Hash}{Hash}
\SetKwFunction{seed}{seed}
\SetKwFunction{Return}{return}
\SetKwFunction{ElementScore}{ElementScore}
\KwData{threshold $\tau$, stream of elements of the form (key,subkey), where keys are from
  domain ${\cal X}$}
\KwOut{set of pairs $(x,c_x)$ where $x\in {\cal X}$}
$\Counters \gets \emptyset$  \tcp*[h]{Initialize a cache of distinct counters}\;
\ForEach(\tcp*[h]{Process a stream element}){stream element with key $x$ and subkey $b$}
{
\If{$x$ is in $\Counters$}{$\Counters{x}.\Merge{x,y}$;}
\Else{
 \If(\tcp*[h]{Create $\Counters{x}$}){$\Hash{x,y} < \tau$}{$\Counters{x}.\Initialize$\; $\Counters{x}.\Merge{x,y}$}
}
}
\Return{For $x \in \Counters$, $(x,\Counters{x}.\CardEst)$}
}
\end{algorithm2e}
}

\subsection{Fixed-size distinct weighted sampling} \label{fixedcache:sec}

  The fixed-size Distinct Weighted Sampling (\dWSnoSp) algorithm is specified for a cache size
  $k$.
Compared with the fixed-threshold algorithm, we keep
some additional state for each cached
key:
\begin{itemize}
\item
The threshold $\tau_x$ when $x$ entered the cache (represented in the
pseudocode as $\Counters{x}.\tau$).  The purpose of maintaining
$\tau_x$  is deriving
confidence intervals on $w_x$. Intuitively, $\tau_x$ captures a prefix of elements
with key $x$ which were seen before the distinct structure for $x$ was
initialized, and is used to estimate the number of distinct subkeys in
this prefix.
\item
A value $\seed{x} \equiv \min_{(x,y) \text{in stream}} \Hash{x,y}$
which is the minimum $\Hash{x,y}$ of all elements with key $x$.
(in the pseudocode, $\Counters{x}.\seed$ represents $\seed{x}$).   Note that it
suffices to track $\seed{x}$ only after the key $x$ is inserted into the cache,
since all elements that occurred before the key entered the cache
necessarily had $\Hash{x,y} > \tau_x$, as the entry threshold $\tau$ can only decrease over time.
\end{itemize}

  The fixed-size \dWS algorithm retains in the cache only the
  $k$ keys with lowest seeds.
The effective threshold value $\tau$ that we work with
is the seed of the most recently evicted key.   The effective  threshold
has the same role as the fixed threshold since it determines the
(conditional) probability on inclusion in the sample for a key with
certain $w_x$. A pseudo code is provided as Algorithm \ref{distinctSHk:alg}.

\begin{algorithm2e}[h]
\caption{{\small Fixed-size streaming Distinct Weighted Sampling (\dWSnoSp)}\label{distinctSHk:alg}}
\DontPrintSemicolon
{\small
\SetKwFunction{Merge}{Merge}
\SetKwFunction{Initialize}{Init}
\SetKwFunction{CardEst}{CardEst}
\SetKwFunction{rand}{rand}
\SetKwFunction{Hash}{Hash}
\SetKwFunction{Return}{return}
\SetKwFunction{ElementScore}{ElementScore}
\KwData{cache size $k$, stream of elements of the form (key,subkey), where keys are from
  domain ${\cal X}$}
\KwOut{set of $(x,c_x,\tau_x)$ where $x\in {\cal X}$}
$\Counters \gets \emptyset$; $\tau \gets 1$  \tcp*[h]{Initialize a cache of distinct counters}\;
\ForEach(\tcp*[h]{Process a stream element}){stream element with key $x$ and subkey $b$}
{
  \If{$x$ is in $\Counters$}{$\Counters{x}.\Merge{x,y}$\;
    $\Counters{x}.\seed \gets \min\{\Counters{x}.\seed,\Hash{x,y}\}$}
  \Else{
    \If(\tcp*[h]{Create $\Counters{x}$}){ $\Hash{x,y} < \tau$}{
      $\Counters{x}.\Initialize$\;
      $\Counters{x}.\Merge{x,y}$\;
      $\Counters{x}.\seed \gets \Hash{x,y}$\;
      $\Counters{x}.\tau \gets \tau$\;
      \If{$|\Counters|>k$}{
        $x \gets \arg\max_{y\in \Counters} \Counters{y}.\seed$\;
        $\tau \gets \Counters{x}.\seed$\;
        Delete \Counters{x}\;
      }
    }
  }
}
\Return{For $x$ in $\Counters$,
  $(x,\Counters{x}.\CardEst,\Counters{x}.\tau)$}
}
\end{algorithm2e}

\subsection{Analysis and estimates} \label{analysis:sec}
 We first consider the sample distribution $S$ of \dWSnoSp.  As
we mentioned (Section \ref{SHsubsec}), it is known
 that classic \SH\ applied with weights $h_x$ has the property that the
 set of sampled keys  is a ppswor sample according to $h_x$ \cite{flowsketch:JCSS2014}.
Surprisingly, the sample distribution properties of \SH\ carries over
  from being with respect to $h_x$ (classic \SH) to being with respect
  to $w_x$ (\dWSnoSp).
We obtain that key $x$ is very likely to be sampled  when
$w_x \gg  \max_{i\in[0,k-1]}(m-\sum_{x\in top_i} w_x)/(k-i)$
where $top_i$ is the set of $i$ heaviest keys.
A detailed explanation of this bound with relevant proofs are omitted due to lack of space and can be found in the technical report~\cite{CompletePaper}.

\subsubsection{Estimate quality and confidence interval}\label{dwsAnalysis}

With the fixed-threshold scheme, we
expect the sample size to include $\tau \sum_y w_y $ keys even when all keys
have $w_x=1$.  With the fixed-size (\dWSnoSp) scheme, we expect the cache to
include keys with $w_x
\gg \sum_y w_y /k$ but it may also include some keys with
small weight.

 For many applications, an
 estimate on the weight $w_x$ of the heavy hitters is needed.
  We compute an estimate with a confidence interval on
  $w_x$ for each cached key $x$, using the entry threshold $\tau$
(or $\Counters{x}.\tau$ in the fixed-size scheme)
and the approximate distinct count $\Counters{x}.\CardEst$.

We obtain the confidence interval
$[\Counters{x}.\CardEst - a_\delta \sigma_2, \Counters{x}.\CardEst-1+1/\tau + a_\delta \sqrt{\sigma_1^2+\sigma_2^2 }]$

where $a_\delta$ is the coefficient for confidence $1-\delta$
according to the normal approximation.  E.g., for 95\% confidence we
can use $a_\delta=2$.

We note the
  confidence intervals are tighter (and thus better)  for keys that are presented earlier and thus have
  $\tau_x \ll \tau$.

Further explanations can be found in the technical report~\cite{CompletePaper}.

\subsection{\idWS design} \label{integrate:sec}
\SetKwFunction{BucketOf}{BucketOf}

 We propose a seamless design (\idWSnoSp) which integrates the
 hashing performed for the weighted sampling component with the hashing performed for
 the approximate distinct counters.  We use a particular type of
 distinct counters based on
{\em stochastic averaging} ($\ell$-partition)
 \cite{FlajoletMartin85,hyperloglog:2007}  (see \cite{ECohenADS:TKDE2015} for an
 overview). This design hashes strings to $\ell$ buckets
and maintains the minimum hash in each bucket.
These counters are the industry's choice as they use
 fewer hash computations.
We estimate the distinct counts using
the  tighter  HIP estimators \cite{ECohenADS:TKDE2015}.
Pseudocode for the fixed-size \idWS is provided as
Algorithm \ref{distinctSHkint:alg}. The parameter $k$ is the sample
size and the parameter $\ell$ is the number of buckets.
Note, we use two independent random hash functions applied to strings:
$\BucketOf$ returns an integer $\sim
U[0,\ell-1]$ selected uniformly at random.
$\Hash$ returns $\sim U[0,1]$ ($O(\log m)$ bits suffice).

 As in the generic Algorithm~\ref{distinctSHk:alg}, we maintain an
object $\Counters{x}$ for each cached key $x$.
The object includes the entry threshold $\Counters{x}.\tau$ and $\Counters{x}.\seed$,
which is the minimum  $\Hash{x,y}$ of all elements $(x,y)$ with key
$x$.
The object also  maintains $\ell$ values $c[i]$ for
 $i=0,\ldots,\ell-1$ from the range of $\Hash$, where
$c[i]$ is the minimum $\Hash$ over all elements $(x,y)$ such that
the element was processed after $x$ was cached and $\BucketOf{x,y}$  is equal to $i$ ($c[i]=1$
when this set is empty). Note that
\newline
$\Counters{x}.\seed \equiv \min_{i\in [0,\ell-1]} c[i]$.
The object also maintains a HIP estimate $\CardEst$ of the number of distinct
subkeys since the counter was created.

\begin{algorithm2e}[h]
\caption{\idWS  \label{distinctSHkint:alg}}
\DontPrintSemicolon
{\small
\SetKwFunction{Merge}{Merge}
\SetKwFunction{Initialize}{Init}
\SetKwFunction{CardEst}{CardEst}
\SetKwFunction{rand}{rand}
\SetKwFunction{Hash}{Hash}
\SetKwFunction{Return}{return}
\SetKwFunction{ElementScore}{ElementScore}
\KwData{cache size $k$, distinct structure parameter $\ell$, stream
  of (key,subkey) pairs}
\KwOut{set of $(x,c_x,\tau_x)$ where $x\in {\cal X}$}
$\Counters \gets \emptyset$; $\tau \gets 1$  \tcp*[h]{Initialize a cache of distinct counters}\;
\ForEach(\tcp*[h]{Process a stream element}){stream element with key $x$ and subkey $y$}
{
  \If{$x$ is in $\Counters$}{
    \If{$\Hash{x,y} < \Counters{x}.c[\BucketOf{x,y}]$}{
      $\Counters{x}.\CardEst \overset{+}{\gets} \ell/\sum_{i=0}^{\ell-1} \Counters{x}.c[i]$\;
      $\Counters{x}.c[\BucketOf{x,y}] \gets \Hash{x,y}$\;
      $\Counters{x}.\seed \gets \min\{\Counters{x}.\seed,\Hash{x,y}\}$\;
    }
  }
  \Else{
    \If(\tcp*[h]{Initialize $\Counters{x}$}){ $\Hash{x,y} < \tau$}{
      \lFor{$i=0,\ldots,\ell-1$}{$\Counters{x}.c[i]\gets 1$}
      $\Counters{x}.\CardEst \gets 0$\;
      $\Counters{x}.c[\BucketOf{x,y}] \gets \Hash{x,y}$\;
      $\Counters{x}.\seed \gets \Hash{x,y}$\;
      $\Counters{x}.\tau \gets \tau$\;
      \If{$|\Counters|>k$}{
        $x \gets \arg\max_{y\in \Counters} \Counters{y}.\seed$\;
        $\tau \gets \Counters{x}.\seed$\;
        Delete $\Counters{x}$\;
      }
    }
  }
}
\Return{For $x\in \Counters$,
  $(x,\Counters{x}.\CardEst,\Counters{x}.\tau)$}
}
\end{algorithm2e}

 For a sampled $x$, we can obtain a
confidence interval on $w_x$ using the lower end point
$\Counters{x}.\CardEst+1$, with error controlled by the distinct
counter
and the upper end point
\newline
$\Counters{x}.\CardEst +
1/\Counters{x}.\tau$, with error controlled by both the distinct
counter and the entry threshold.   The errors are combined as explained in
Section~\ref{dwsAnalysis} using the HIP error of
 $\sigma_2 \approx (2\ell)^{-0.5} \Counters{x}.\CardEst\ .$

 The size of our structure is
$O(k\ell \log m)$  and the representation of the $k$ cached keys.
Note that the parameter
 $\ell$ can be a constant for DDoS applications:  A choice of $\ell=50$
gives NRMSE of 10\%.
Note, that this design can be further optimized according to resource constraints as explained in the technical report~\cite{CompletePaper}.

\ignore{
 We can further optimize this design according to the
most constrained resources in our application.  Let it be
processing time, memory, or maximum processing time per element.
For example, to control element processing time
we can evict more keys (a fraction of the cache) when it is full.
When memory is highly constrained we can instead use the
exponent representation (round $c[i]$ to an integral power of $2$) as done
with Hyperloglog \cite{hyperloglog:2007}  and apply an appropriate HIP
estimate as described in \cite{ECohenADS:TKDE2015}.  This will reduce
the structure size to $O(k\log m + k\ell \log\log m)$.
}

\section{Combined Weighted Sampling}\label{combo:sec}

 We now present our \cWS algorithm for combined heavy hitters detection.
 The pseudocode, which builds on
our \idWS design (Algorithm \ref{distinctSHkint:alg}),
 is presented in  Algorithm \ref{comboSHk:alg} and works with
a specified parameter $\rho$.
 For each cached key $x$, the combined weighted sampling (\cWSnoSp) algorithm
also includes a classic counter $\Counters{x}.f$  of the number of
elements with key $x$ processed after $x$ entered the
 cache.

\begin{algorithm2e}[h]
\caption{Streaming \cWS  \label{comboSHk:alg}}
\DontPrintSemicolon
%
{\small
\SetKwFunction{Merge}{Merge}
\SetKwFunction{Initialize}{Init}
\SetKwFunction{CardEst}{CardEst}
\SetKwFunction{rand}{rand}
\SetKwFunction{Hash}{Hash}
\SetKwFunction{Return}{return}
\SetKwFunction{ElementScore}{ElementScore}
\KwData{cache size $k$, distinct structure parameter $\ell$, parameter $\rho$,
  stream (key,subkey) pairs}
\KwOut{set of $(x,c_x,f_x, \tau_x)$ where $x\in {\cal X}$}
$\Counters \gets \emptyset$; $\tau \gets 1$  \tcp*[h]{Initialize a cache of distinct counters}\;
\ForEach(\tcp*[h]{Process a stream element}){stream element with key $x$ and subkey $y$}
{  $erand \gets 1-(1-\rand{})^{1/\rho}$ \tcp*[h]{Randomization for  $h_x$ count}\;
   \If{$x$ is in $\Counters$}{
    $\Counters{x}.f \overset{+}{\gets} 1$\tcp*[h]{Increment count}\;
    \If{$\Hash{x,y} < \Counters{x}.c[\BucketOf{x,y}]$}{
      $\Counters{x}.\CardEst \overset{+}{\gets} \ell/\sum_{i=0}^{\ell-1} \Counters{x}.c[i]$\;
      $\Counters{x}.c[\BucketOf{x,y}] \gets \Hash{x,y}$\;
      $\Counters{x}.\seed \gets
      \min\{\Counters{x}.\seed,\Hash{x,y},erand \}$\;
    }
  }
  \Else{
    \If(\tcp*[h]{Initialize $\Counters{x}$}){ $\min\{erand,\Hash{x,y}\} < \tau$}{
      \lFor{$i=0,\ldots,\ell-1$}{$\Counters{x}.c[i]\gets 1$}
      $\Counters{x}.\CardEst \gets 0$\;
      $\Counters{x}.f \gets 1$\;
      $\Counters{x}.c[\BucketOf{x,y}] \gets \Hash{x,y}$\;
      $\Counters{x}.\seed \gets \min\{\Hash{x,y},erand\}$\;
      $\Counters{x}.\tau \gets \tau$\;
      \If{$|\Counters|>k$}{
        $x \gets \arg\max_{y\in \Counters} \Counters{y}.\seed$\;
        $\tau \gets \Counters{x}.\seed$\;
        Delete $\Counters{x}$\;
      }
    }
  }
}
\Return{For $x \in \Counters$,
  $(x,\Counters{x}.\CardEst,\Counters{x}.f,\Counters{x}.\tau)$}
}
\end{algorithm2e}

 Similarly to \dWSnoSp,  if we are only interested in the set of
 sampled keys (cHH candidates), it suffices to maintain the seed values
 of cached keys without the counting and distinct counting structures.
The counters are useful for obtaining
estimates and confidence intervals on the  combined weights of cached
keys:
For  a desired confidence level $1-\delta$.
  The lower end of the interval is
$\Counters{x}.\CardEst + \rho \Counters{x}.f -a_\delta \sigma_1$,
where $\sigma_1$ is the standard error of the distinct count.
 For the higher end, we bound the contribution of the prefix,
 which has expectation bounded by $1/\tau-1$, and subject both to the
 \SH\ error and the approximate distinct counter error, so we obtain
\newline
$\Counters{x}.\CardEst + \rho \Counters{x}.f -a_\delta \sigma_1
-1+1/\tau + a_\delta\sqrt{\sigma_1^2+\sigma_2^2}$.

%% file: evaluation.tex
\section{Evaluation} \label{eval:sec}

\subsection{Theoretical Comparison}
In Table \ref{table:methods comparison} we show a theoretical memory usage comparison of our algorithms, SuperSpreaders and Locher \cite{Locher:spaa2011}, assuming all algorithms use the same distinct count primitive. We are using the notations in Table \ref{table:symbols}, $\delta$ as the probability that a given source becomes a false negative or a false positive, $N$ as the number of distinct pairs, $r$ as the number of estimates, $s$ as the number of pairs of distinct counting primitives used to compute each estimate, and c (for a $c$-superspreader (i.e. we want to find keys with more than $c$ distinct elements) choosing $c=\tau^{-1}$.
As we can also see from the table, the cache size affects the distinct weight estimation error for the keys.

\ignore{We assume all algorithms use the same distinct count primitive. Using the notations in Table \ref{table:symbols}, $\delta$ as the probability that a given source becomes a false negative or a false positive, $N$ as the number of distinct pairs, $r$ as the number of estimates, $s$ as the number of pairs of distinct counting primitives used to compute each estimate, and c (for a $c$-superspreader (i.e. we want to find keys with more than $c$ distinct elements) choosing $c=\tau^{-1}$, Table \ref{table:methods comparison} shows the theoretical memory usage comparison.
}

\begin{center}
\begin{table}[h]
\small
\begin{tabular}{|m{2.6cm}||m{2.8cm}|m{2.4cm}|}\hline
Algorithm & Memory usage & Keys' distinct weight estimation error  \\ \hhline{|=||=|=|}
Fixed-threshold distinct WS & $O(\tau \sum_y w_y \cdot \ell\log m)$ (Exp.)& $\tau^{-1} + w_y/\sqrt{2\ell}$ \\ \hline
Fixed-size \dWSnoSp & $O(k\ell\log m)$ & $(1/k) \sum_y w_y  +  w_y/\sqrt{2\ell}$ \\ \hline
Superspreaders 1-Level Filtering~\cite{superspreaders:NDSS2005} & $O(\frac{N}{c})$ & $NA$ \\ \hline
Superspreaders 2-Level Filtering~\cite{superspreaders:NDSS2005} & $O(\frac{N}{c} ln\frac{1}{\delta})$ & $NA$ \\ \hline
Locher~\cite{Locher:spaa2011} & $O(rs \cdot 2\ell+|k|)$ & $NA$ \\ \hline
\end{tabular}
 \caption{Theoretic Comparison between methods}
    \label{table:methods comparison}
    \end{table}
\end{center}



%
%
%
%
\subsection{Fixed-size distinct weighted sampling (\dWSnoSp)}

We have done extensive testing of our algorithms both on real internet traffic traces and on synthetically generated data.

We begin our evaluation with the results of tests done on packets from a trace from The CAIDA UCSD Anonymized Internet Traces 2014~\cite{CAIDA_Equinox_2014}. For each packet, the destination IP address is the key, and the source IP address is the value. In order to display the full ability of our algorithm, we added into this data synthetic packets which form keys with many distinct values. Specifically, the synthetic packets all have unique subkeys and contain $4$ keys, such that the keys have cardinalities $2000$, $1000$, $500$ and $250$.
The entire data is made up of approximately $1M$ ($993750$) $\{key, value\}$ pairs, containing $33977$ keys and $52859$ distinct pairs.
\begin{figure*}
       \centering
       \begin{subfigure}[b]{0.31\textwidth}
                \includegraphics[width=\linewidth,height=1.6in,clip=true,trim=0mm 0mm 0mm 0mm, scale=0.2]{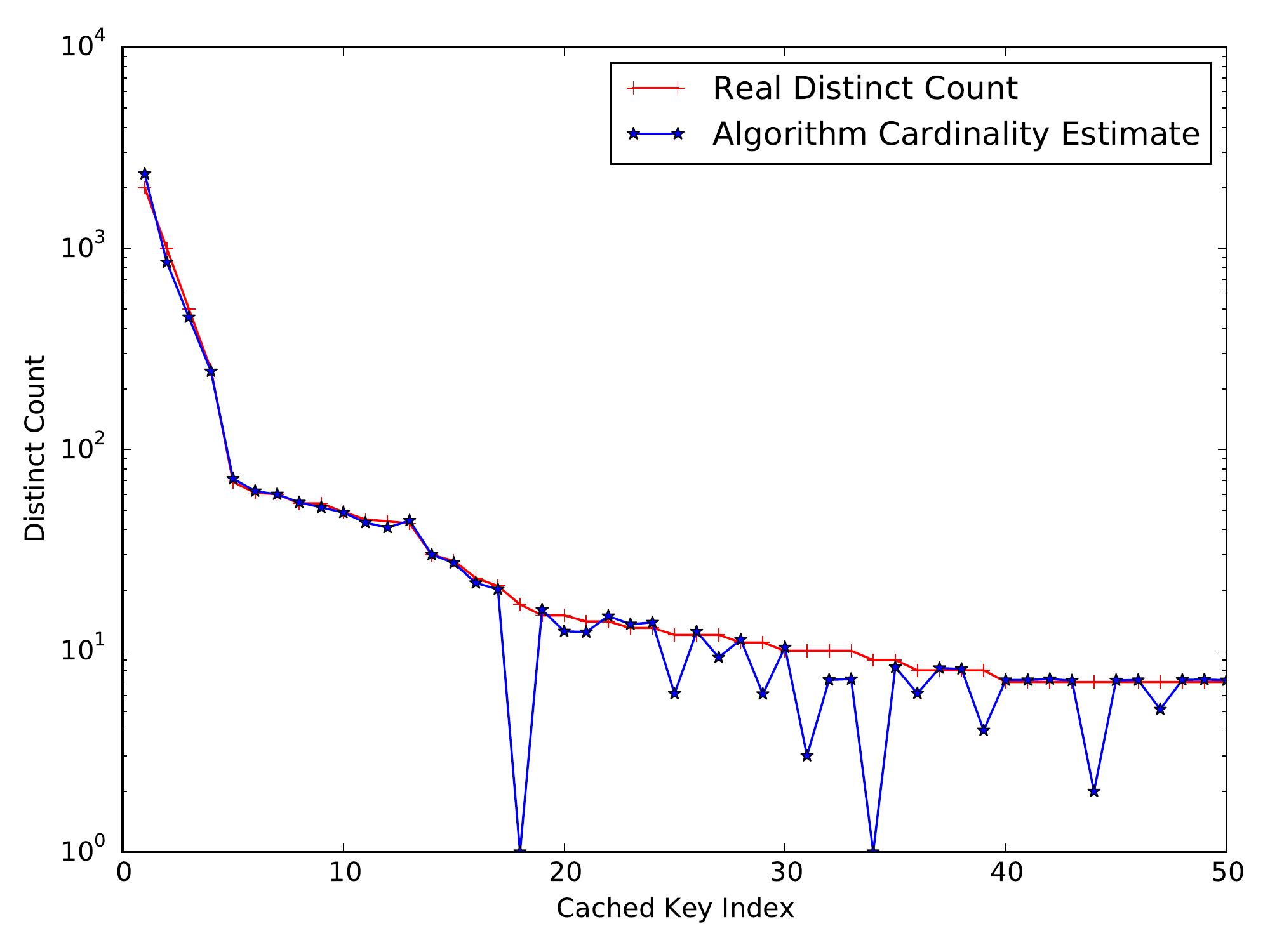}
                \caption{Cardest vs the real distinct count.}
                \label{Figure:UpdatedDhhAccuracy2}
        \end{subfigure}
        \hfill
        \begin{subfigure}[b]{0.31\textwidth}
                \includegraphics[width=\linewidth,height=1.7in,clip=true,trim=00mm 00mm 00mm 00mm, scale=0.2]{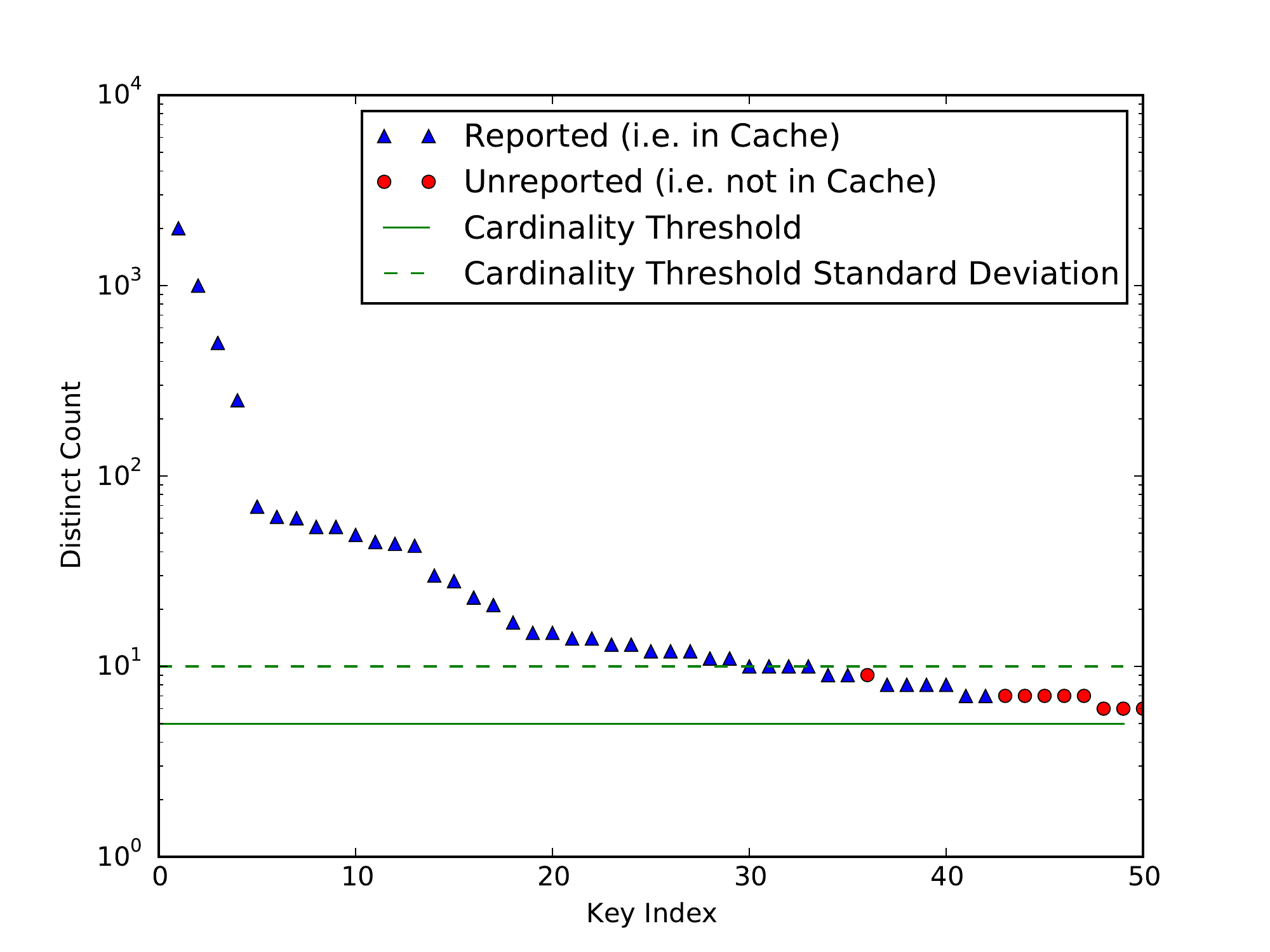}
                \caption{Distinct count reported vs. unreported.}
                \label{Figure:UpdatedDhhFN3}
        \end{subfigure}
        \hfill
        \begin{subfigure}[b]{0.32\textwidth}
                \includegraphics[width=\linewidth,height=1.5in,clip=true,trim =0mm 0mm 0mm 0mm, scale=0.2]{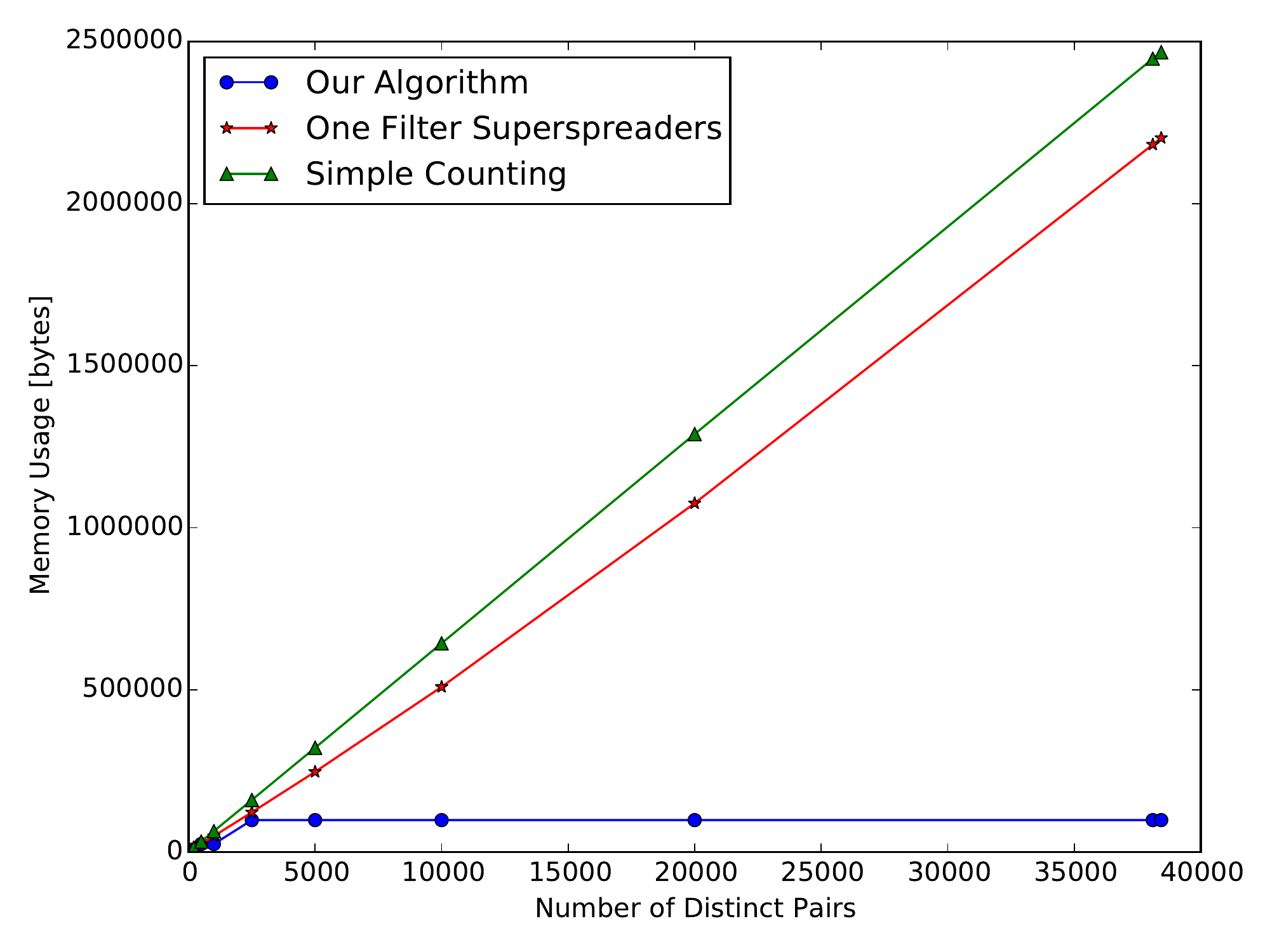}
                \caption{Memory usage for distinct pair count.}
                \label{Figure:NewUpdatedDhhSpace}
        \end{subfigure}
        \caption{Distinct Weighted Sampling (\dWSnoSp) test results (cache 2000)}\label{fig:DHHTests}
\end{figure*}

\begin{figure*}
       \centering
       \begin{subfigure}[b]{0.32\textwidth}
                \includegraphics[width=\linewidth,clip=true,trim=0mm 0mm 0mm 0mm, scale=0.2]{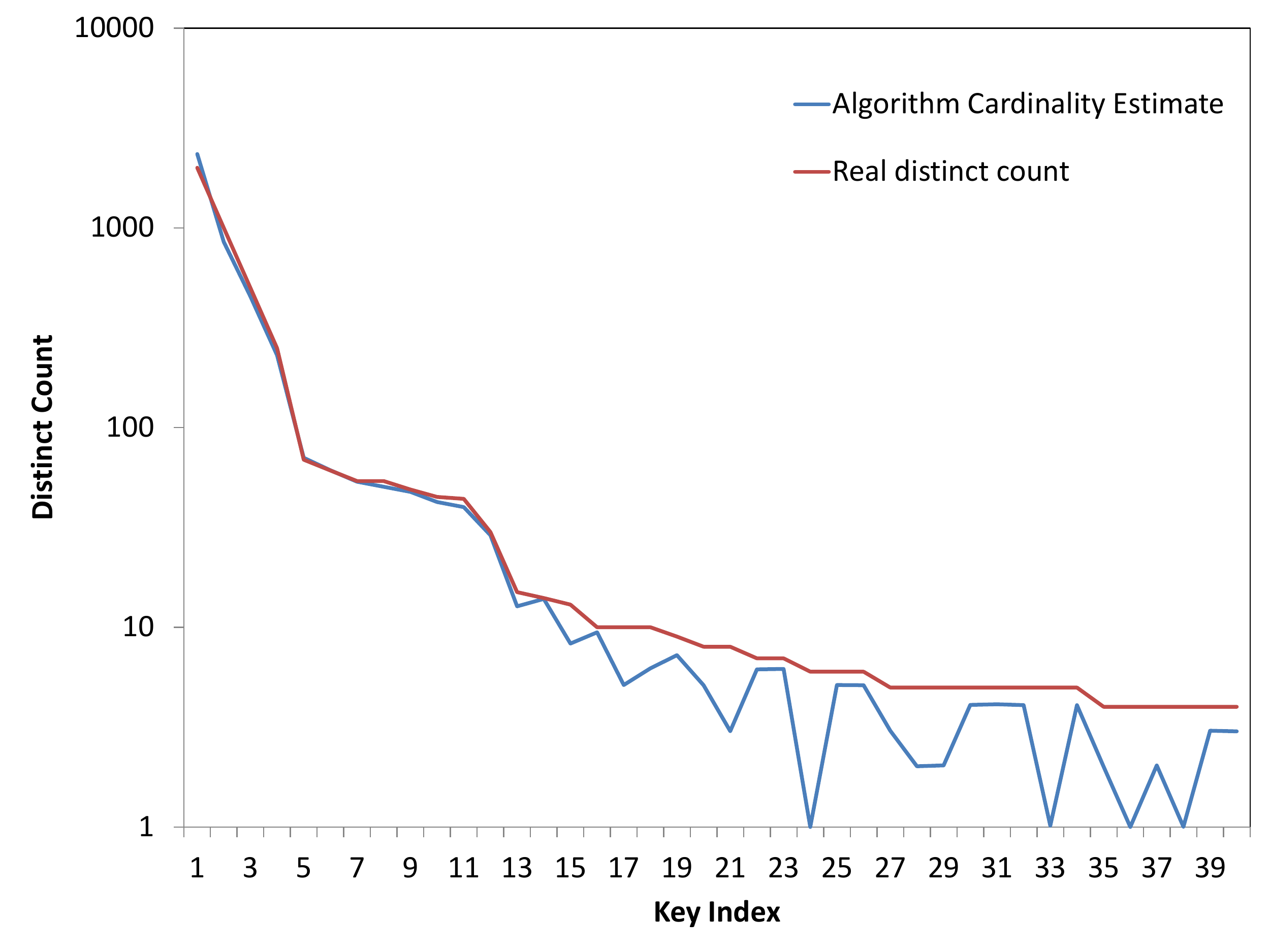}
                \caption{Cardest vs the real distinct count.}
                \label{Figure:UpdatedDhhAccuracy}
        \end{subfigure}
        \hfill
        \begin{subfigure}[b]{0.32\textwidth}
                \includegraphics[width=\linewidth,clip=true,trim=0mm 0mm 0mm 0mm, scale=0.2]{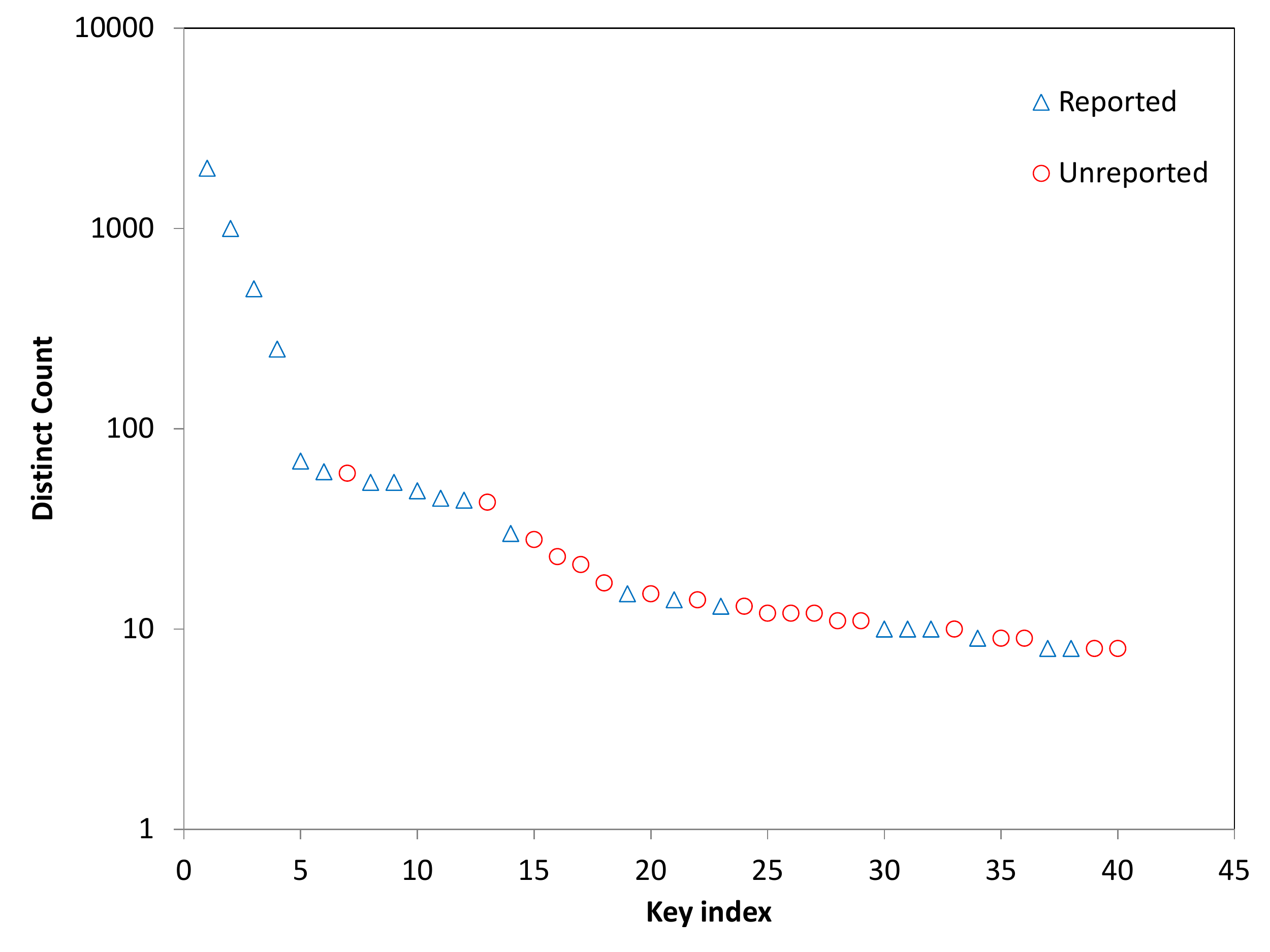}
                \caption{Distinct count reported vs. unreported.}
                \label{Figure:UpdatedDhhFN}
        \end{subfigure}
        \hfill
        \begin{subfigure}[b]{0.32\textwidth}
                \includegraphics[width=\linewidth,clip=true,trim =0mm 0mm 0mm 0mm, scale=0.2]{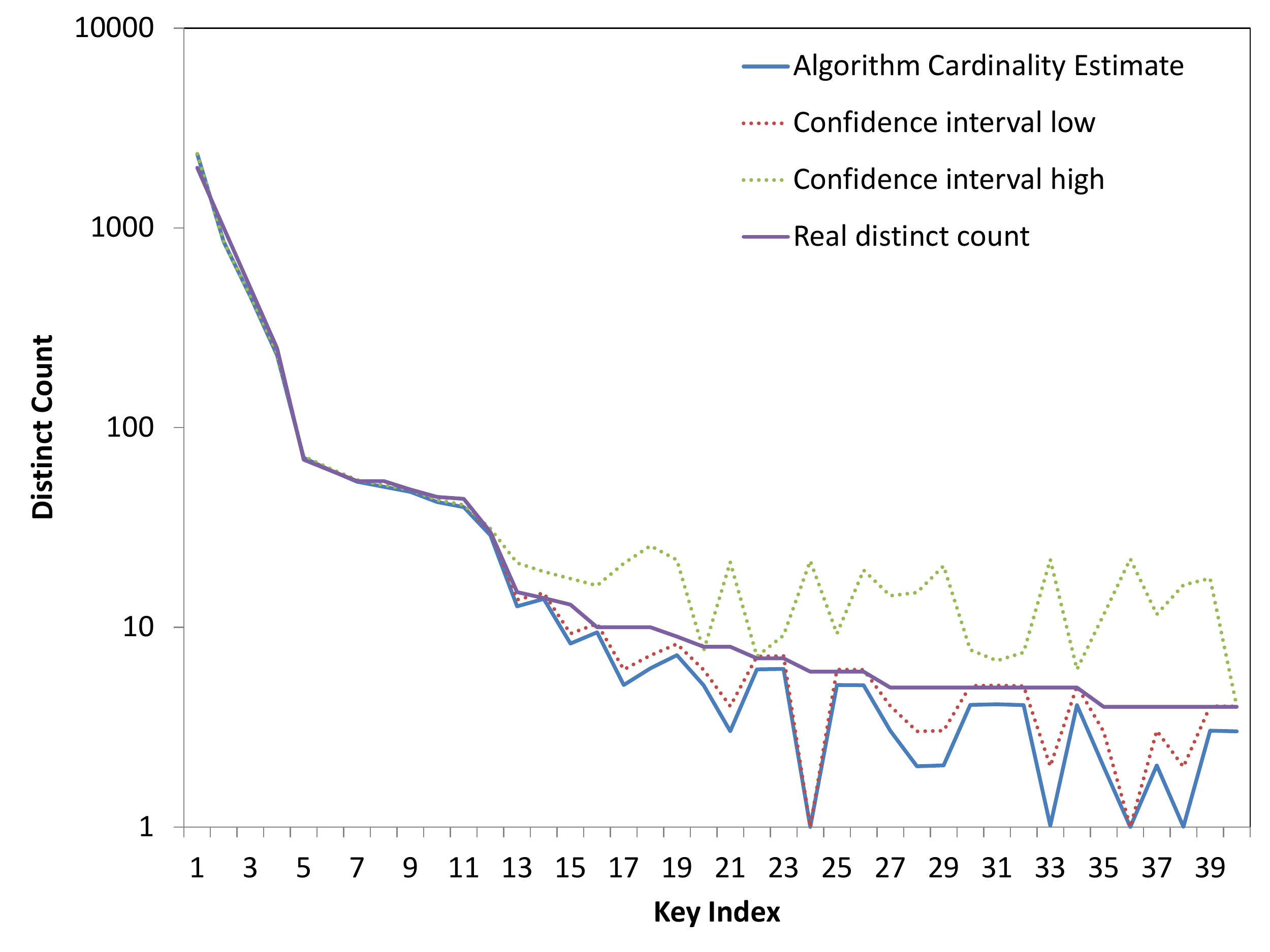}
                \caption{Accuracy with Confidence interval}
                \label{Figure:UpdatedDhhSpaceConf}
        \end{subfigure}
        \caption{Distinct Weighted Sampling (\dWSnoSp) test results (cache 500)}\label{fig:DHHTests2}
\end{figure*}

We evaluate the accuracy of our Fixed-Size \dWS algorithm (see Section~\ref{fixedcache:sec}). The results presented are based on the implementation of Algorithm~\ref{distinctSHkint:alg}. We set the cache size $k=2000$ and distinct structure parameter $\ell=64$.
We compare, both in terms of accuracy and space, our algorithm to a simple and highly inefficient algorithm which counts the number of distinct values associated with each key.
Fig.~\ref{Figure:UpdatedDhhAccuracy2} shows the cardinality estimate
(Cardest) calculated by the algorithm in this test, compared to the
actual distinct count of each of these keys. We can see that the algorithm provides relatively accurate estimates, with the average error of the Cardest, measured by the average distance between the Cardest and the real distinct count was $1.5$, with a median error of $1$. In terms of memory usage, our algorithm consumes a constant amount of space, while the simple algorithm consumes space that is linear with the number of distinct pairs seen. The memory consumption of both algorithms is depicted in Fig.~\ref{Figure:NewUpdatedDhhSpace}. While we chose to compare memory usage with the simplified Superspreaders algorithm - the one filter algorithm, its two filter variant algorithm reaches a better asymptotic memory usage model. However, the two filter variant is more complicated and its memory usage is more susceptible to implementation factors but in any event, its memory usage still grows linearly with the stream length. 
Fig.~\ref{fig:DHHTests2} shows the injected Caida tests performed with cache size $k=500$. The confidence interval for the cardinality estimate is shown in Fig. \ref{Figure:UpdatedDhhSpaceConf}
As explained in Section~\ref{analysis:sec}, our algorithm seeks to find keys with a cardinality over $t= \sum_y w_y/k$. 
To evaluate the false positive (FP) and false negative (FN) rates of our algorithm, we set t to be the threshold and evaluate the keys' deviation. Any key $x$ with a cardinality $w_x>t$ that is not reported by our algorithm (not in cache) and has a high deviation from t (normalized by the cardinality error), is a FN. 
FPs are defined respectively - a key $x$ that is reported with a cardinality $w_x<t$ and a high deviation from t. Fig.~\ref{Figure:UpdatedDhhFN3} shows the distinct count of keys that were reported (in cache) by the algorithm, and those that were not reported (not in cache - the red colored circles shown). 
Note that generally, the algorithm will also cache keys which have a distinct count much
lower than $t$.  We use the Cardest to select only keys with a Cardest
that is above $t$.
\ignore{In the example shown, only $12$ out of the $2000$
cached keys had Cardest value higher than the real distinct count, and
as stated, these overestimates were within one standard deviation. As a result,
we obtained a near zero FP rate.}
In the example shown, only $12$ out of the $2000$
cached keys had Cardest value higher than the real distinct count, all of these overestimates were within the confidence interval bounds - within error bounds. In the same example, we found 18 (9 depicted in graph) potential FN - unreported keys above the threshold set, but all of which are within one standard deviation shown by the dotted line.  Note that we can tune parameters to obtain a near-zero FN rate instead of a near-zero FP since 
both the number of overall distinct pairs as well as the cache size affect this threshold.

\subsection{Combined Weighted Sampling}
\begin{figure}[h]
\begin{center}
\includegraphics[trim = 0mm 0mm 0mm 0mm, scale=0.35]{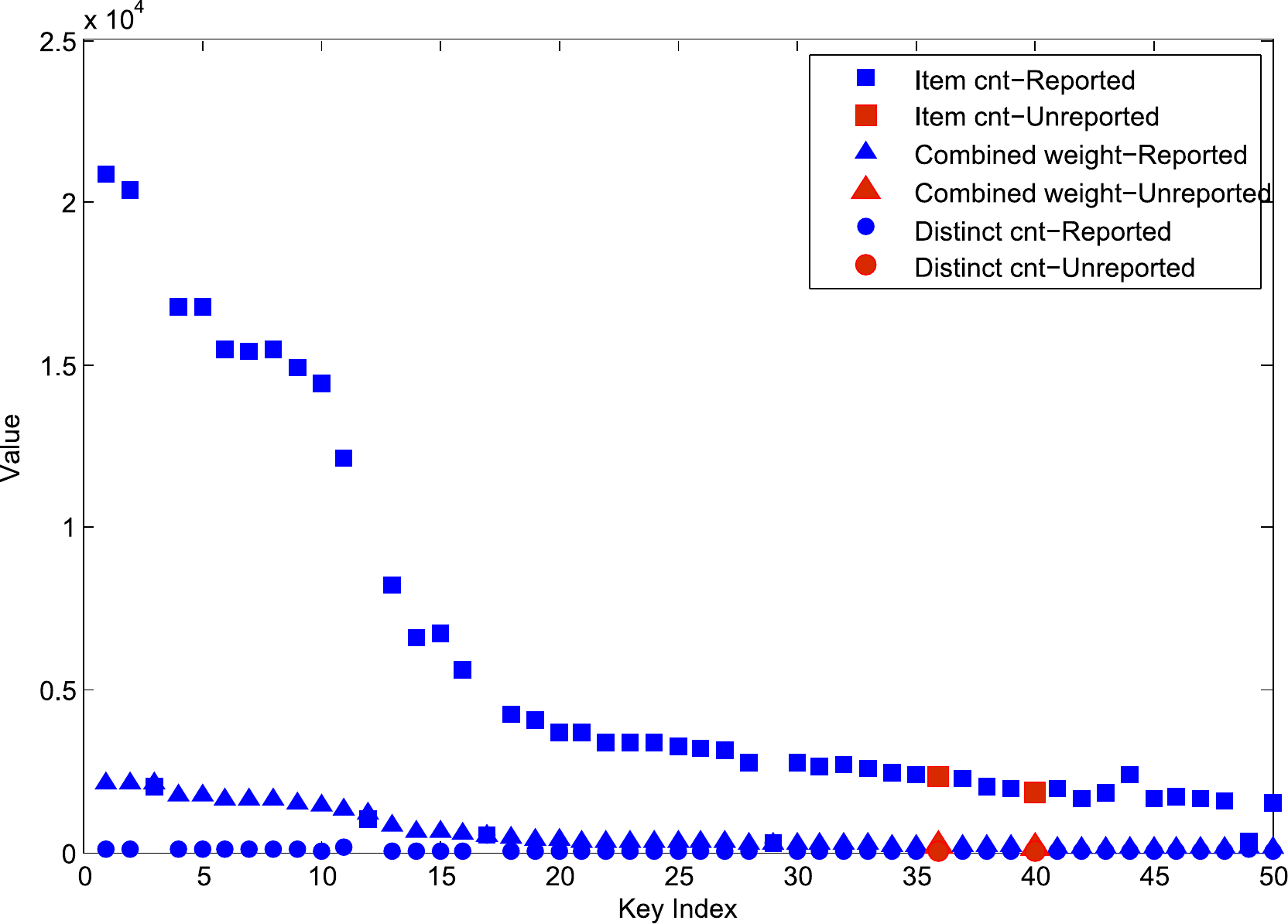}
\end{center}
\caption{\cWSnoSp: Algorithm estimates of reported vs. unreported keys.}
\label{Figure:UpdatedChhFN}
\end{figure}

We test our streaming \cWS algorithm (see Section~\ref{combo:sec}) on the data described above, with $\rho=0.1$.
The \cWS algorithm samples keys according to their combined
weight.  This is depicted in Fig.~\ref{Figure:UpdatedChhFN}, where for
each key we compare the actual count, the distinct count and the
combined weight of the key. We can see that
the algorithm properly identifies the keys with highest combined
weights and that those keys are different than when
sampling by distinct weights.
For example, item $28$ which is unreported in
Fig.~\ref{Figure:UpdatedDhhFN}, has a very high number of non-distinct
queries and therefore is reported as item $18$ in
Fig.~\ref{Figure:UpdatedChhFN}.

%% file: newDNS.tex

\section{Randomized Sub-Domain DDoS Attacks on DNS}\label{dns:sec}


\subsection{Attack Description}
 The DNS is a hierarchical distributed naming system for translating more readily domain names to the numerical IP addresses. The DNS distributes the responsibility by designating authoritative name servers for each domain.


The Random Subdomain DDoS (RSDDoS) attack on DNS (also known as the Random QNAME attack or the Nonsense Name attack~\cite{NonsenseName:Liu2015}) has recently become a rising threat to the DNS service.
In this type of attack, the queries sent for the victim's domain include many different randomly generated or highly varying sub-domains.
For example, an attack query may be of the form: \emph{bjsufyd.www.google.com: type A, Class IN}

%
%
%
While the main target might be the victim's authoritative name
server, the attack also has collateral damage and impacts the Open
Resolver domain name servers as well as DNS-caches, which are
operated by intermediate Internet service providers. This occurs
since these queries will avoid the caching mechanism of the open
resolver.
%
\ignore{
\begin{figure}[h]
\begin{center}
\includegraphics[trim = 1mm 1mm 1mm 60mm, clip, scale=0.2]{DNSAttack}
\end{center}
\caption{DNS Random Sub-Domain attack overview}
\label{Figure:dnsOverview}
\end{figure}
}

Detection of RSDDoS attacks is difficult due both to the large
number of sources from which the queries are received, as well as
the general structure of the attack packets. The packets that
comprise this type of attack are DNS queries which have a legitimate
form but
request a nonexistant domain or subdomain. 
Detection 
is made even more difficult by the
increasing usage of disposable domains. These are large volumes of
automatically generated domains, legitimately created by top sites
and services (e.g., social networks and search engines), to give some
signal to their server~\cite{DisposableDomains:ChenAPNDL14}.

The current solution of internet providers so far has been to manually identify the targeted zone and to temporarily prevent the name server from sending queries for this zone~\cite{NonsenseName:Liu2015} (or alternatively to reduce the number of queries using rate limiting).
The solution we propose 
identifies the targeted zone automatically. Furthermore, by analyzing the peacetime traffic we are able to automatically identify some of the legitimate requests for the targeted domain, to significantly reduce the false positives of our system. Using our proposed system, attacks can be mitigated quickly and accurately.

\subsection{Attack Query Identification System}\label{dnsSystem:sec} 


We provide an overview for a system which identifies attack queries of the form {\em VAR.victimdomain.com}. That is, queries that consist of a random (or automatically generated) string as a prefix of the domain (in the least significant domain sub-part), which have so far been the most common query form in these attacks. The system detects RSDDoS attacks on DNS servers and creates signatures for subsequent mitigation.
Queries are processed with the key being the {\em victimdomain.com} and the subkey being the {\em VAR} part of the query.
We are currently developing a system 
that expands mitigation capabilities to additional forms of these attacks. 
This advanced system is beyond the scope of this paper.

Traffic analysis is done in two stages. The first stage is a preprocessing of peacetime traffic, the second is an analysis of traffic during an attack.
Our analysis makes use of both a \emph{Distinct Heavy Hitters} module such as the combined sketch described in Section~\ref{combo:sec} and a \emph{Classic Heavy Hitters} module such as that of Metwally et. al.~\cite{spacesaving:ICDT2005}.


\begin{figure*}
       \centering
       \begin{subfigure}[b]{0.49\textwidth}
                \includegraphics[trim = 0mm 0mm 0mm 0mm, clip, scale=0.23]{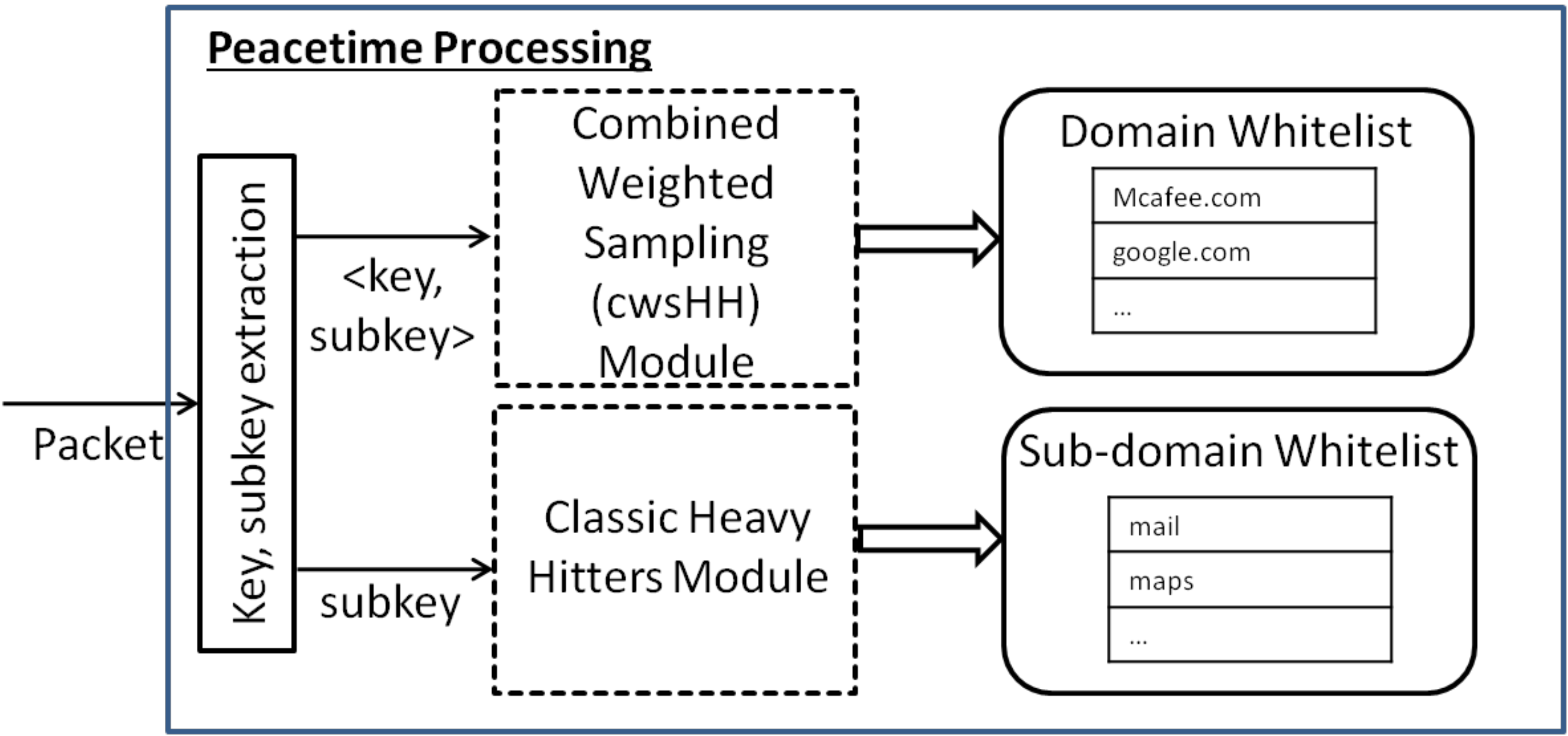}
                \caption{Peacetime packet processing}
                \label{Figure:peacetimeOverview}
        \end{subfigure}
        \hfill
        \begin{subfigure}[b]{0.49\textwidth}
                \includegraphics[trim = 0mm 0mm 0mm 0mm, clip, scale=0.23]{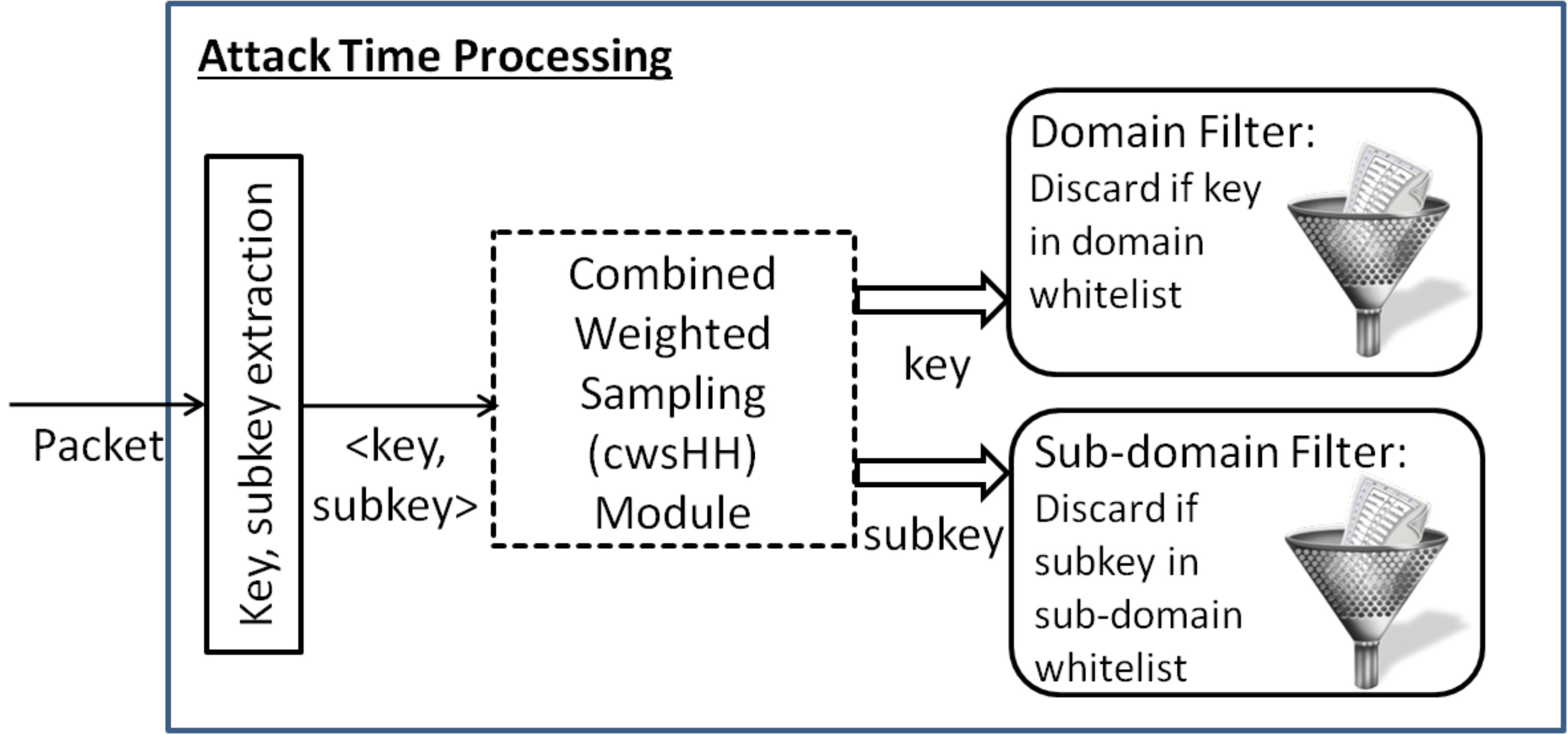}
                \caption{Attack time packet processing}
                \label{Figure:attackOverview}
        \end{subfigure}
        \caption{Packet Processing}\label{fig:DNSOverview}
\end{figure*}

\ignore{
\begin{figure}[h]
\begin{center}
\includegraphics[trim = 0mm 0mm 0mm 0mm, clip, scale=0.25]{DNSSystemPeace}
\end{center}
\caption{Packet Processing}
\label{Figure:peacetimeOverview}
\end{figure}
}
As can be seen in Fig.~\ref{Figure:peacetimeOverview}, during peacetime, queries are processed as follows:
\begin{enumerate}[wide, labelwidth=!, labelindent=0pt]

\item Query parsing: $\langle key, subkey \rangle$ is extracted.

\item The $\langle key, subkey \rangle$ pair is inserted into our streaming combined weighted sampling module: 
This mechanism identifies zones (keys) that are heavily queried, and have a large number of distinct subdomains (subkeys). The output of this module is used to create a white-list of zones that are combined heavy hitters and most likely use disposable domains as part of their routine behaviour. 

\item The subkey is inserted into a classic heavy hitters module: The output of this module generates a white-list of subdomains (subkeys) that appear frequently in the DNS queries.   When mitigating an attack, the requests with a subkey from this list should not be blocked. The motivation for this is to identify strings which are commonly used as subdomains, and are therefore not likely to be randomly generated. Using this white-list minimizes the false positives when mitigating an attack with the signatures we generate.  
\end{enumerate}


\ignore{
\begin{figure}[h]
\begin{center}
\includegraphics[trim = 0mm 0mm 0mm 0mm, clip, scale=0.25]{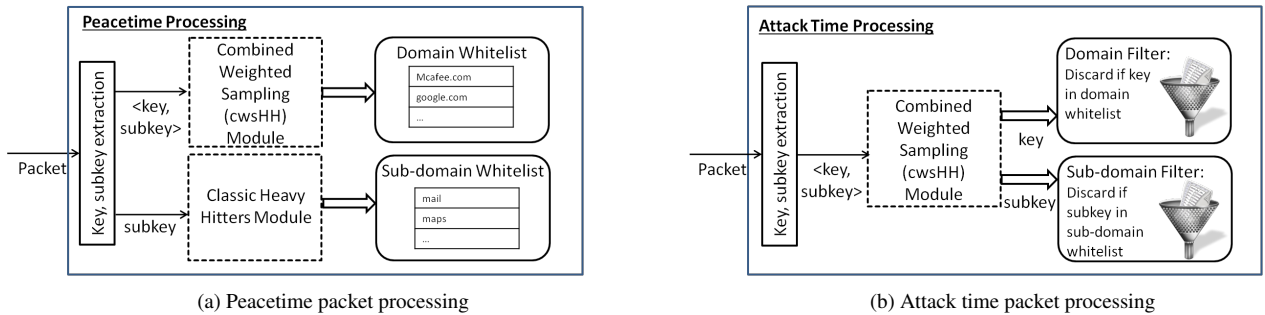}
\end{center}
\caption{Packet Processing}
\label{Figure:attackOverview}
\end{figure}
}
As shown in Fig.~\ref{Figure:attackOverview}, in the detection phase the system identifies the attacked domains as follows: 
\begin{enumerate}[wide, labelwidth=!, labelindent=0pt]

\item Query parsing: key and subkey are extracted.

\item  The $\langle key, subkey \rangle$ pair is inserted into our streaming combined weighted sampling module: During an RSDDoS attack, the
zones which are identified as having highly distinct subdomains and
are heavy hitters, are suspected as being the victims. The output of this module is therefore used to
create an initial set of attack signatures.

\item White-list filter: The white-lists created during the peacetime processing are used to filter out legitimate queries and therefore reduce the false positive rate of the system. Two types of filters are used: 
    \begin{itemize}
    \item \emph{Subkey white-list:} If the query subkey was identified as being frequent during peacetime, we assume that this is not an automatically generated subkey and therefore the query is considered to be legitimate. In this manner, legitimate queries of attacked domains may be serviced.

    \item \emph{Key white-list:} the zones that have been identified
    as white-list domains in peacetime are
    filtered out since these are zones which have a large number of distinct sub-domains as part of their regular operation, e.g., disposable domains.
    We note however, that detection of attacks on
    disposable domains (which have been white-listed) is left for the full paper where
    the amounts of distinct values detected by the combined weighted sampling algorithm at attack time are compared
    with those of peace time, and if a notable increase is observed then an attack is
    signaled and a corresponding (see below) mitigation is suggested.
    Note that bloom filters can be used to speed up whitelist
    search as described in~\cite{BDHH15}.
    \end{itemize}
\end{enumerate}

The above system creates a list of domains which are
likely under attack. Mitigation is thus proposed by filtering out DNS
requests to these domains except if the {\em VAR} part appears
in the subdomain white-list generated at peacetime. Therefore,
the suggested mitigation would first allow the queries for domains with white-list subkeys,
and then block the requests to the domains suspected to be under attack.

\subsection{System Evaluation}

The RSDDoS DNS attack query identification system presented above has been evaluated on traces of actual RSDDoS attacks captured by a large ISP.

We analyzed $5$ captures which were sniffed during different RSDDoS attacks
and contained both attack and legitimate DNS queries. All captures
were taken within a single month in $2014$. Note that most of the
captures contain $5000$ queries as that was the set amount that was
sniffed for each attack spotted. The ISP identified the
RSDDoS attacks as they were occurring. The victim was  the
authoritative name servers of the ISP (the victim domain belongs to
one  of the ISP's clients) or the Open Resolvers of the ISP (the
victim domain does not belong to the ISP).  We compare our results
to the analysis performed manually by the ISP. We use Cache size
$k=50$ and distinct structure parameter $\ell=256$. Note that, some
of the attacks analyzed had a very high percentage of distinct
queries and others had lower rates.
The repetitions are of randomly generated queries that were each
repeated several times in the traffic. These different attack rates
are an example for the usefulness of both the cHH and dHH algorithms.
As we did not have access to a peacetime capture to obtain a
whitelist, we generated a whitelist based on domains with a
relatively high distinct subdomain count.

Consider attack $1$ in Table~\ref{table:DNSDetection}. The capture
consisted of $92469$ DNS queries. Of these, $4133$ are attack
queries targeted at the same zone, with a randomly generated least
significant domain sub-part, containing $2051$ distinct queries,
meaning that some of the queries were repeated. Of the $4133$
queries, the system counted $4123$. Meaning, that $10$ queries for
the attacked zone had gone through before the zone was placed in the
structure (i.e., in the cache). Once inside, the zone was not
evicted from the structure at any point, all subsequent queries were
counted, hence $99.8\%$ of the queries were identified.
\begin{center}
\begin{table}[h]
\small
\begin{tabular}{|p{0.8cm}||p{1.5cm}|p{1.0cm}|p{1.8cm}|p{1.9cm}|}\hline
Source & Queries in capture  & Attack queries & Distinct attack queries & Attack queries identified  \\ \hhline{|=||=|=|=|=|}
1 & $92469$ & $4133$ & $2051$ & $99.8\%$ \\ 
2 & $5000$& $389$ & $367$ & $99.7\%$ \\ 
3 & $5000$ & $602$ & $567$ & $100\%$ \\ 
4 & $5000$& $334$ & $330$ & $100\%$ \\ 
5 & $5000$ & $3364$ & $631$ & $99.8\%$ \\ \hline 
\end{tabular}
 \caption{Results on Real DNS Attack Captures}
    \label{table:DNSDetection}
    \end{table}
\end{center}

%% file: applications.tex

\section{Additional Applications}\label{applications:sec}

Our algorithms are instrumentals in the detection and mitigation of
various DDoS and other network attacks. In this section we provide
several examples of such attacks.

\subsection{SYN Attacks}

A classic form of DDoS attacks still common in today's networks is
the TCP SYN Attack. In this form of attack, the attacker initiates
many TCP connections, while never completing the TCP handshake. The
connection queue of the target is therefore filled up with
incomplete connections, preventing it from addressing new connection
requests from legitimate parties. The attacker may make an attack more
difficult to detect by utilizing a botnet or a large army of sources
for carrying out the attack or even by simply using spoofed sources.
In this case, the attacked destination receives connection
requests from many different sources. Using our
algorithm, we can identify destinations which have a large number of
distinct sources, and thus be able to identify the attack soon after
the number of requested connections exceeds normal use.

\subsection{Email Spam}

Email spam is often characterized by having a single sender or a
small number of senders which send emails to a large number of
different recipients.  By making the sender the key and the
addressee as the subkey, our algorithm can easily detect spammers.
Using peacetime along the lines suggested in Section \ref{dns:sec}
can improve the system accuracy and reducing false positives.

\subsection{Flash Crowd Detection}

Combined heavy hitters may be used, for example, when network load is a concern.
Consider a flash event. A flash crowd or flash event, is a situation
where a very large number of users simultaneously access some web
site~\cite{FlashvsDDoS}. For example, a major developing news event
may cause a flash crowd in major news web sites. Fast and automated
detection and processing (mitigation or other responses) of such
anomalies is important for maintaining robustness of the network
service.

To identify the affects a flash crowd has on the entire network, it
is not sufficient to solely identify the rise in the distinct number
of users accessing a site. Instead, we would like to identify that
there are both many accesses to this site causing a high load on the
network as well as many different users who are accessing this site.
For a network administrator to reallocate network resources to meet
this demand, both of these measurements are significant. Our \cWS
algorithm can be used to identify the rise in
both parameters.

%% file: relatedWork.tex
\section{Related work} \label{related:sec}

The concept of distinct heavy hitters, together with the motivation
for DDoS attack detection,  was introduced
  in a seminal paper of
Venkataraman et al \cite{superspreaders:NDSS2005}.
 Their algorithm, aimed at detection of fixed-threshold heavy hitters,
returns as
candidate heavy hitters the keys with an (initialized) Bloom filter
that is filled beyond some threshold.
Keys with high count in the sample are likely to be heavy hitter and
almost saturate their bloom filter.
A related work  adapts dHH schemes to TCAMs
\cite{Bandi:ICDCS2007}.
  Our fixed-threshold scheme is conceptually related to
  \cite{superspreaders:NDSS2005}.  Some key differences are the
better tradeoffs we obtain by using
 approximate distinct counters instead of Bloom filters, and our
simpler structure with analysis that ties it directly
to classic analysis of weighted sampling, which also simplifies the
use of parameters.
  More importantly,  we
provide a solution to the fixed-size problem and also address
the estimation problem.  The estimates
on the weight of the heavy keys that can be obtained from the Bloom
filters in \cite{superspreaders:NDSS2005} are much weaker, since once the filter is saturated,
it can not distinguish between heavy and very
 heavy keys.

 Locher  \cite{Locher:spaa2011} recently presented two designs for dHH
 detection which makes use of approximate distinct counters.
 The first design is sampling-based and builds on the distinct pair sampling
 approach of \cite{superspreaders:NDSS2005}.  This design also only
 applies to the fixed-threshold problem.  The other design uses
linear sketches and applies to the fixed-size problem.
Locher's designs are weaker than ours both in terms of practicality and
in terms of theoretical bounds.
The linear-sketch based design
utilizes linear-sketch based distinct counters, which are much less efficient in
practice that the sampling-based ones.
The designs have a
quadratically worse  dependence of structure size on the detection
 threshold $\tau$, which is
$\Omega(\tau^{-2})$ instead of our $O(\tau^{-1})$.
Finally, multiple copies of the
 same structure are maintained to boost up confidence, which results in
 a large overhead, since heavy hitters are accounted for in
most copies.
Locher's  code was not available for a direct comparison.

  Another conceivable approach is to convert to DHH classic  fixed-size deterministic HH streaming
  algorithms,
such as Misra Gries \cite{MisraGries:1982} or
the space saving
algorithm \cite{spacesaving:ICDT2005},
by replacing counters with
approximate distinct counters.
 The difficulty that arises is that the same distinct element may
 affect the structure multiple times when the same key re-enters the cache, resulting in much weaker
 guarantees on the quality of the results.